\documentclass[prb,twocolumn,preprintnumbers,superscriptaddress]{revtex4}
\usepackage{graphicx}
\usepackage{dcolumn}
\usepackage{bm}
\usepackage{graphicx}  
\usepackage{dcolumn}   
\usepackage{bm}        
\usepackage{verbatim}   
\usepackage[usenames]{color}  
\usepackage{stmaryrd}
\usepackage{float}
\usepackage{units}
\usepackage{textcomp}
\usepackage{textcomp}
\usepackage{amsmath}
\usepackage{commath}
\usepackage{amssymb}
\usepackage{esint}
\usepackage{ae,aecompl}
\usepackage[colorlinks=true,linkcolor=blue,citecolor=blue]{hyperref}
\begin{document}

\title{Rashba-splitting of the Dirac points and the symmetry breaking in the strained artificial graphene}

\author{Vram Mughnetsyan}

\affiliation{Department of Solid State Physics, Yerevan State University, Alex Manoogian 1, 0025 Yerevan, Armenia}

\author{Aram Manaselyan}
\affiliation{Department of Solid State Physics, Yerevan State University, Alex Manoogian 1, 0025 Yerevan, Armenia}

\author{Manuk Barseghyan}
\affiliation{Department of Solid State Physics, Yerevan State University, Alex Manoogian 1, 0025 Yerevan, Armenia}

\author{Albert Kirakosyan}
\affiliation{Department of Solid State Physics, Yerevan State University, Alex Manoogian 1, 0025 Yerevan, Armenia}

\author{David Laroze}
\email{dlarozen@uta.cl}
\affiliation{Instituto de Alta Investigaci\'{o}n, CEDENNA, Universidad de Tarapac\'{a},
Casilla 7D, Arica, Chile}

\begin{abstract}
The effect of Rashba spin-orbit interaction and anisotropic elastic strain on the electronic, optical and thermodynamic properties of artificial graphene-like superlattice composed of InAs/GaAs quantum dots has been considered theoretically. The electronic energy dispersions have been obtained using Green's function formalism in combination with the Fourier transformation to the reciprocal space and an exact diagonalization technique. We have observed a splitting of Dirac points and appearance of additional Dirac-like points due to the Rashba spin-orbit interaction. Breaking of the hexagonal symmetry of the dispersion surfaces caused by the strain anisotropy is observed as well. It is shown that both the spin-orbit interaction and strain anisotropy have a qualitative impact on the measurable characteristics of the considered structure and can be used as effective tools to control the performance of devices based on artificial graphene.
\end{abstract}

\maketitle

\maketitle

\section{Introduction}
Dirac materials are foreseen to be of paramount importance because of their universal behavior and the robustness of their properties which are linked to symmetry \cite{Wehling,Polini}. Their band structure is similar to one of relativistic massless particles  where the energy dependence on the momentum is linear in the vicinity of touching (Dirac) points of electronic bands. Graphene is an innate example of one-atom-thick two-dimensional electron system composed of carbon atoms on a honeycomb lattice with two inequivalent sites in the unit cell. Due to its unique electronic spectrum, graphene makes possible the observation and test of table-top quantum relativistic phenomena in experiments, which are unobservable in high-energy physics \cite{Geim}.

In principle Dirac-type singularities can exist in any 2D lattice with the same underlying symmetry as in graphene. The advanced methods such as atom-by-atom assembling \cite{Gomes}, nanopatterning of two-dimensional electron gas in semiconductors \cite{Singha}, and optical trapping of ultracold atoms in crystals of light \cite{Tarruell} make possible to design and fabricate artificial honeycomb lattices or artificial graphene (AG) which is a unique structure for the realization, investigation, and manipulation of a wide class of systems displaying massless Dirac quasiparticles, topological phases, and strong correlations. One of the reasons for pursuing the study of AG is the opportunity to reach regimes in these systems which are difficult to achieve in graphene, such as high magnetic fluxes, tunable lattice constants, and precise manipulation of defects, edges, and strain \cite{Ferrari}. These can enable tests of several predictions for massless Dirac fermions.

It has been shown that two-dimensional electron gas in a periodic potential of the honeycomb array of GaAs/GaAlAs quantum dots (QD) can result in isolated massless Dirac points with controlable Fermi velocity \cite{Gibertini}. The controllable Fermi velocity in its turn can lead to bound states of Dirac fermions \cite{Downing}, which is crucial for building practical digital devices with a well-defined on/off logical state \cite{Yung}.
 The realization of massless Dirac fermions in conventional semiconductors opens an interesting scenario related to the impact of spin-orbit interaction (SOI), particularly if one uses InAs-based materials such as honeycomb lattice of InAs/GaAs QDs.

Although the growth of homogeneous and spatially ordered arrays of InAs/GaAs QDs remains a technological challenge \cite{Costantini,Inoue}, some recent works point to the possibility of controlling the size, shape, as well as the electron concentration in them using strain engineering and selective area epitaxy. In this regard there is a good prospect to achieve uniform, position-controlled InAs QDs in the near future \cite{Mohan,Maier,Alonzo-Gonzalez,Cheng,Liang,Carmesin}.

It is known that the elastic strains at the InAs/GaAs heterojunction due to the lattice mismatch dramatically alters the electronic band structure \cite{Goerbig, Asano, Lim, Tadic}. It has been shown that the strain anisotropy in InAs/GaAs AG leads to the shift of the Dirac points from the $K$ and the $K^{\prime}$ points of the first Brillouin Zone (FBZ) resulting to anisotropy in the Fermi velocity and qualitative changes in the density of states (DOS) \cite{Mughnetsyan1}.

The optical properties of transistors \cite{Schwierz}, optical switches \cite{Gan,Yao,Liu1}, midinfrared photodetectors \cite{Sun1,Freitag}, photovoltaic devices \cite{Liu2}, ultrafast lasers \cite{Sun2}, etc., significantly depend on the light-matter interaction, which is limited in graphene (optical absorption is less than $2.5 \%$). One of the advantages of AG is the possibility to overcome this limitation and tune the absorption coefficient (AC) by means of external factors such as Rashba SOI. The possibility to study collective optical response of modulated nearly-2D electrons \cite{Maksym,Peeters} and holes \cite{Sarkisyan} in semiconductors is another advantage of AGs based on QDs.

The heat capacity (HC) as a measurable thermodynamic quantity can be considered as a sensitive tool to bring out the modifications in the electron energy spectrum in QD as well as in graphene structures due to the internal and external factors \cite{Ma, Maksym, Oh, Castano-Yepes}. The study of HC in AG is of grate interest due to possibility to observe combine effects originated from the quantum confinement in QDs and the underlying honeycomb symmetry.

In this regard the consideration of the Rashba SOI and the elastic strain field in AG opens new perspectives for the control of the optical and thermal properties of Dirac fermions.

In the present work the effect of Rashba SOI on the electronic band structure, DOS as well as optical and thermodynamic properties of AG composed of highly strained InAs/GaAs QDs has been considered. The manuscript is organized as follows: In Sec. II the model and the method are presented. In Sec. III the results are displayed and the corresponding discussion is given. The conclusions are presented in Sec. IV.

\section{Theoretical model}
Our theoretical model is based on the following assumptions. In the  view of strong quantization in the direction perpendicular to the plane of the superlattice (SL) we will assume that electron makes two-dimensional motion in the plane of SL. Further, due to very weak dependence of the hydrostatic strain on the coordinate in the transverse direction, only the in-plane variations of the strain will influence on the motion of electron \cite{Mughnetsyan2}.

The method developed in Ref. \cite{Andreev} allows one to derive an analytic expression for the Fourier components of the strain tensor for a single QD of arbitrary shape in a material with lattice of cubic symmetry (see Appendix: A). In the framework of the mentioned approach the hydrostatic strain in a two-dimensional SL of honeycomb symmetry, composed of cylindrical QDs of the height $h_{d}$ and the radius $r_{d}$ is as follows \cite{Mughnetsyan1}:
\begin{widetext}

\begin{equation}
\label{HS}
\begin{aligned}
&\tilde{\varepsilon}_{h}(\xi_{1},\xi_{2})=\sum\limits_{i=1}\limits^{3}\int d\xi_{3} \tilde{\varepsilon}_{ii}(\xi_{1},\xi_{2},\xi_{3})=
\varepsilon_{0}\tilde{\chi}_{QD}(n_{1},n_{2}) \Bigg(3- \frac{C_{11}+2C_{12}}{\pi}
\int \limits_{-\infty}\limits^{\infty}\frac{\xi^{-1}_{3}\sin(\xi_{3}h_{d}/2)d\xi_{3}}
{C_{12}+C_{44}+\left(\Lambda_\xi \right)^{-1}}\Bigg),
\end{aligned}
\end{equation}
\end{widetext}
where
\begin{equation}
\Lambda_\xi=\sum\limits_{p=1}\limits^{3}\frac{\xi_{p}^{2}}{C_{44}\xi^{2}+C_{an}\xi_{p}^{2}}
\end{equation}
and
\begin{equation}
\label{SF}
\begin{aligned}
& \tilde{\chi}_{QD}(n_{1},n_{2})=\frac{2\pi r_{d}J_{1}(r_{d}|\vec{G}|)}{s_{0}|\vec{G}|}A(n_{1},n_{2})(1-A(n_{1},n_{2})),
\end{aligned}
\end{equation}
is the Fourier component of the SL's shape function \cite{Gunawan,Mughnetsyan2}, $J_{1}(t)$ is the first kind Bessel function of the first order, $\vec{G}=n_{1}\vec{g}_{1}+n_{2}\vec{g}_{2}$ is the 2D lattice vector in reciprocal space,
$\vec{g}_{1}=(2\pi/3a)(1;\sqrt{3})$ and $\vec{g}_{2}=(2\pi/3a)(1;-\sqrt{3})$ are the elementary vectors of the reciprocal lattice, $a$ is the smallest distance between the centers of QDs in the SL, $A(n_{1},n_{2})=\exp{(-i2\pi(n_{1}+n_{2})/3)}$, $n_{1,2}$ are integers, $s_{0}$ is the area of the SL's unite cell, $\tilde{\varepsilon}_{ii}(\vec{\xi}\hspace{0.1cm})$ is the 3D Fourier transform of the diagonal element of the strain tensor in SL, $\xi_{1}=G_{x}$, $\xi_{2}=G_{y}$, $\xi^{2}=\sum_{i=1}^{3} \xi_{i}^{2}$, $C_{11}$, $C_{12}$, and $C_{44}$ are the elastic moduli of the matrix material (GaAs),
$C_{an}=C_{11}-C_{12}-2C_{44}$ is the parameter of anisotropy, $\varepsilon_{0}=(a_{1}-a_{2})/a_{2}$ is the initial strain \cite{BookStrain}, $a_{1}$ and $a_{2}$ are the lattice constants of the GaAs and InAs lattices respectively.

It should be noted that  when the condition $h_{d} \ll r_{d}$ is satisfied the dependence of hydrostatic strain on ``$z$" coordinate is weak \cite{Mughnetsyan2}, and Eq.(\ref{HS}) can be used for calculation of the hydrostatic strain in 2D space:
\begin{equation}
\label{2DHS}
\begin{aligned}
& \varepsilon_{h}(\vec{r})=\sum_{\vec{G}} \tilde{\varepsilon}_{h}(\vec{G})e^{i\vec{G}\vec{r}}.
\end{aligned}
\end{equation}
The Hamiltonian of the considered system is
\begin{equation}
\label{Ham}
\mathcal{H}=\frac{1}{2}\hat{p}\frac{1}{m(\vec{r})}\hat{p}
+\mathcal{H}_{SO}+V(\vec{r}),
\end{equation}
were
\begin{equation}
\label{SO}
\mathcal{H}_{SO}=\frac{\alpha}{\hbar}(\vec{\sigma}\times\vec{p})_{z}
\end{equation}
is the Rashba SOI Hamiltonian \cite{Bychkov}, $\sigma_x$,
$\sigma_y$ and $\sigma_z$ are the Pauli matrices, $\alpha$ is the Rashba parameter,
$V(\vec{r})=v_{0}(\vec{r})+a_{c}\varepsilon_{h}(\vec{r})$ is the periodic potential of QD SL, $v_{0}(\vec{r})=Q(E_{g,GaAs}-E_{g,InAs})$ is the potential of unstrained structure, $E_{g,GaAs(InAs)}$ is the band gap of GaAs(InAs) material, $Q$ is the conduction band offset, $a_{c}$ is the hydrostatic potential constant and $m(\vec{r})$ is the electron effective mass.
Due to the periodicity of the Hamiltonian (\ref{Ham}) one can make
a Fourier transformation to the momentum space \cite{Gunawan,MughManSL}:
\begin{equation}
\label{WFFT}
\psi_{\uparrow (\downarrow)}(\vec{r})=\frac{1}{S}e^{i\vec{k}\vec{r}}u_{\vec{k}\uparrow
(\downarrow)}(\vec{r})=\frac{1}{S}\sum_{\vec{G}}u_{\vec{k},
\vec{G}\uparrow (\downarrow)} e^{i (\vec{k}+\vec{G})\vec{r}},
\end{equation}
\begin{equation}
V(\vec{r})=\sum_{\vec{G}}V_{\vec{G}} e^{i \vec{G} \vec{r}},
\end{equation}
\begin{equation}
\label{VMFT}
\frac{1}{m(\vec{r})}=\sum_{\vec{G}}m^{-1}_{\vec{G}} e^{i \vec{G}\vec{r}}.
\end{equation}
Note that in Eq. (\ref{WFFT}) $u_{\vec{k}\uparrow (\downarrow)}(\vec{r})$ and
$u_{\vec{k},\vec{G}\uparrow (\downarrow)}$ are the Bloch amplitude and its Fourier
transform for the spin-up (spin-down) component of the spinor
$\widehat{\psi}$, respectively.  Also, $V_{\vec{G}}$ and $m^{-1}_{\vec{G}}$ are the Fourier
transforms of SL potential and inverse effective mass, respectively. Finally, $\vec{k}$ is
quasi-momentum and $S$ is the effective area of the AG. Substituting the expressions (\ref{WFFT}) -- (\ref{VMFT}) to the Ben
Daniel-Duke's equation $\mathcal{H}\widehat{\psi}=E\widehat{\psi}$
one can arrive to the following set of linear equations in reciprocal space:
\begin{widetext}
\begin{eqnarray}
\label{SysEqA}
\sum_{\vec{G}^{\prime}}\Bigg(\Big[\frac{\hbar^{2}}{2}m^{-1}_{\vec{G}-\vec{G}^{\prime}}(\vec{k}+\vec{G})(\vec{k}+\vec{G}^{\prime})
+V_{\vec{G}-\vec{G}^{\prime}}-E\delta_{\vec{G},\vec{G}^{\prime}}\Big]u_{\vec{k},\vec{G}^{\prime}\uparrow}+\alpha\delta_{\vec{G},\vec{G}^{\prime}}\left[i(k_{x}+G^{\prime}_{x})+ (k_{y}+G^{\prime}_{y})\right]u_{\vec{k},
\vec{G}^{\prime}\downarrow}\Bigg)=0,
\end{eqnarray}
\begin{eqnarray}
\sum_{\vec{G}^{\prime}}\Bigg(\Big[\frac{\hbar^{2}}{2}m^{-1}_{\vec{G}-\vec{G}^{\prime}}(\vec{k}+\vec{G})(\vec{k}+\vec{G}^{\prime})
+V_{\vec{G}-\vec{G}^{\prime}}-E\delta_{\vec{G},\vec{G}^{\prime}}\Big]u_{\vec{k},\vec{G}^{\prime}\downarrow}-\alpha\delta_{\vec{G},\vec{G}^{\prime}}\left[i(k_{x}+G^{\prime}_{x})-(k_{y}+G^{\prime}_{y})\right]u_{\vec{k},\vec{G}^{\prime}\uparrow
}\Bigg)=0,
\label{SysEqB}
\end{eqnarray}
\end{widetext}
where $m^{-1}_{\vec{G}}=\delta_{\vec{G},0}m^{-1}_{GaAs}+(m^{-1}_{InAs}-m^{-1}_{GaAs})\tilde{\chi}_{QD}(\vec{G})$ and $V_{\vec{G}}=(v_{0}/s_{0})\tilde{\chi}_{QD}(\vec{G})$
 are the Fourier transforms of the electron's inverse mass and the SL potential, respectively. The electronic dispersions are obtained by means of diagonalization of the set of equations (\ref{SysEqA}), (\ref{SysEqB}) for each value of the quasi-momentum $\vec{k}$.

The DOS of the considered structure can be expressed as follows:
\begin{equation}
\label{DOS}
\begin{aligned}
& \rho(E)=\frac{1}{(2\pi)^{2}}\sum\limits_{j}\int\limits_{FBZ}\delta(E-E_{j}(\vec{k}))d^{2}k,
\end{aligned}
\end{equation}
where the integration is carried out over the FBZ and $j$ denotes the number of the miniband.

Assuming that the Fermi energy $E_{F}$ is on the touching point between two couples of splitted minibands, the AC caused of the allowed direct transitions is as follows:
\begin{eqnarray}
\label{AC}
\alpha(\omega)=\alpha_{0}\sum_{i=1}^{2}\sum_{j=3}^{4}\int \limits_{FBZ}d^{2}k
|M_{i,j}(\vec{k})|^{2} \nonumber \\
\times \delta(\hbar \omega-(E_{j}(\vec{k})-E_{i}(\vec{k})),
\end{eqnarray}
where
\begin{equation}
\label{MatEl}
\begin{aligned}
M_{i,j}(\vec{k})=\hbar \sum \limits_{\vec{G}} (u^{(i)}_{\vec{k},\vec{G}\uparrow}u^{(j)}_{\vec{k},\vec{G}\uparrow}+u^{(i)}_{\vec{k},\vec{G}\downarrow}u^{(j)}_{\vec{k},\vec{G}\downarrow})(\vec{G}\vec{\eta})
\end{aligned}
\end{equation}
is the dipole matrix element of the transitions from the i-th to the j-th miniband, $\alpha_{0}=e^{2}(m_{0}^{2}c h_{d} \omega \sqrt{\epsilon})^{-1}$,
$\omega$ and $\vec{\eta}$ are the frequency and the vector of polarization of incident photon, $\epsilon$ is the dielectric constant, $m_{0}$ and $e$ are the free electron mass and the charge, respectively and $c$ is speed of light.

We have calculated also the electronic HC of the system using the following expression \cite{Ma,Girifalco}:
\begin{equation}
\label{HeatCap}
    c_V=\int E \rho(E) \frac{\partial f(E,T)}{\partial T}dE,
\end{equation}
were integration is carried out over all the conduction band, $f(E,T)=(e^{\beta(E-\mu(T))}+1)^{-1}$ is the Fermi-Dirac distribution function, $\beta=1/k_BT$ and $\mu(T)$ is the chemical potential.
One can obtain the dependence of the chemical potential on temperature by solving the following equation:
\begin{equation}
\label{EqCP}
    n=\int\rho(E)f(E,T)dE,
\end{equation}
were it is assumed that the electron 2D concentration $n$ in the conduction band is constant and the Fermi energy
$E_F=\mu(T=0)$ is on the touching point between two couples of splitted minibands.

\section{Discussion}
\begin{figure*}
\centerline{\includegraphics[width=15cm]{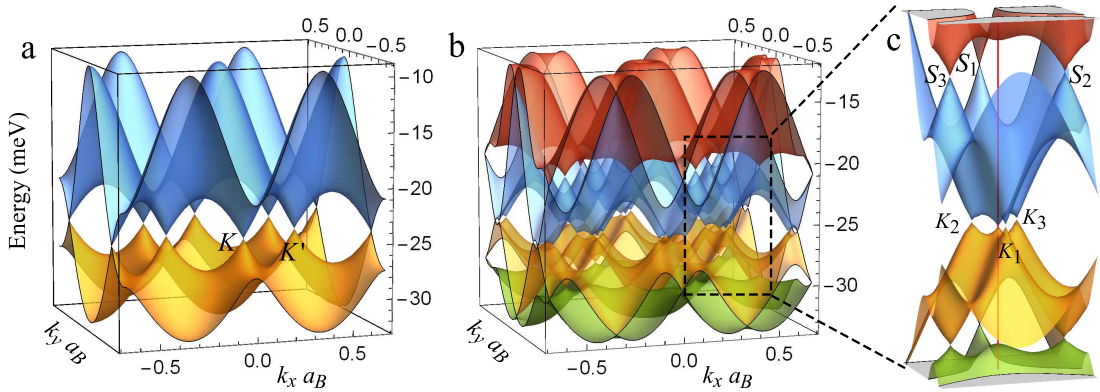}} \caption{(Colour on-line)
Dispersion surfaces for the splitted by SOI electronic minibands of isotropically strained AG. (a) the entire picture, (b) the first four minibands in the vicinity of the $K^{\prime}$ point, (c) two touching minibands in the vicinity of $K^{\prime}$ point (the position of the $K^{\prime}$ point is indicated by a red line).}
\end{figure*}
\begin{figure*}
\centerline{\includegraphics[width=15cm]{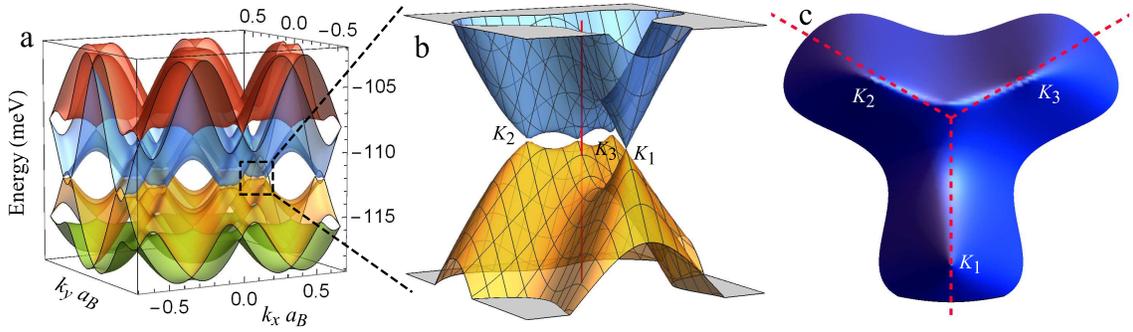}} \caption{(Colour on-line)
(Colour on-line)
Dispersion surfaces for the splitted by SOI electronic minibands of anisotropically strained AG. (a) the entire picture, (b) two touching minibands in the vicinity of $K^{\prime}$ point (the position of the $K^{\prime}$ point is indicated by a red line), (c) the top view of the first miniband dispersion surface (the red dashed lines indicate on the diagonals of three hexagons with the same corner in reciprocal space and their crossing point coincides with the $K^{\prime}$ point.}
\end{figure*}
The numerical calculations are carried out for the following values of parameters: $a=22$nm, $r_{d}=10$nm, $h_{d}=2$nm, $m_{InAs}=0.023m_{0}$, $m_{GaAs}=0.067m_{0}$, $E_{g,GaAs}=1518$meV, $E_{g,InAs}=413$meV and $Q=0.6$ \cite{AdachiHandbook}. Taking into account that the electron is mostly localized in the QD regions, in the absorption coefficient (\ref{AC}) we use the value of the dielectric constant in InAs material ($\epsilon=12.3$). The energy levels broadening are taken into account by replacing the Dirac $\delta$ function in Eq. (\ref{AC}) by Lorentzian function with the value of the broadening parameter $\Gamma=0.2$meV \cite{Arzberger}.
\begin{figure}
\centerline{\includegraphics[width=8cm]{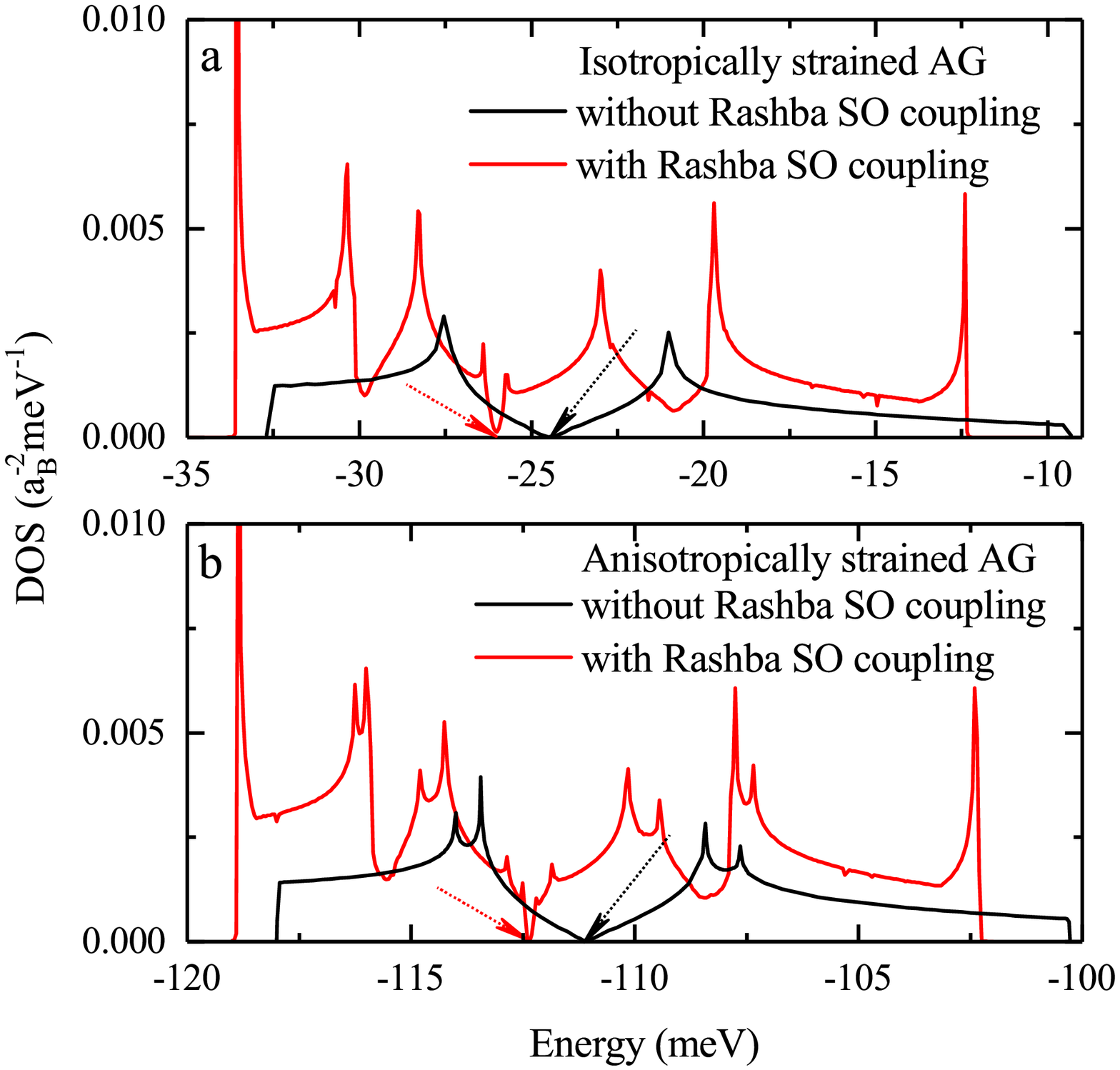}} \caption{(Colour on-line)
Density of states for isotropically (a) and anisotropically (b) strained AG with (red lines) and without (black lines) Rashba SOI. Arrows indicate on the touching points of the first and the second minibands.}
\end{figure}
\begin{figure}
\centerline{\includegraphics[width=8cm]{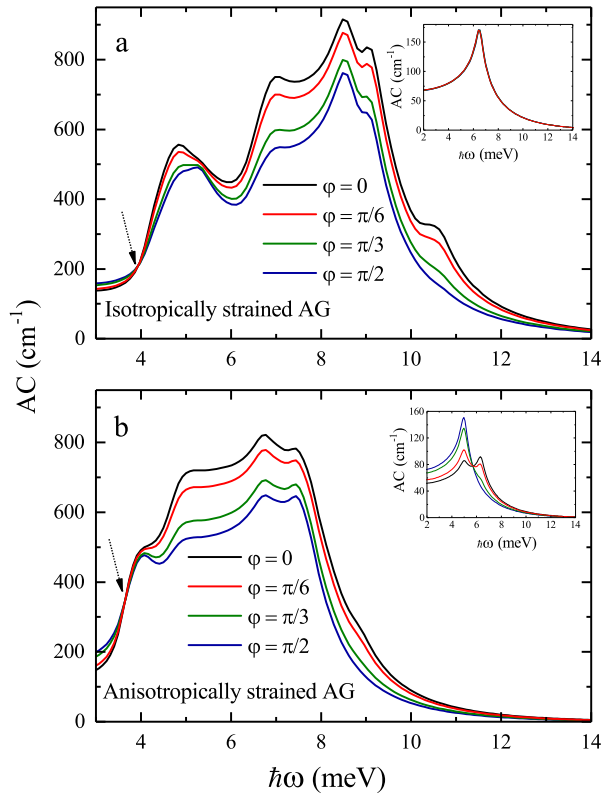}} \caption{(Colour on-line)
Absorption coefficient for isotropically (a) and anisotropically (b) strained AG with Rashba SOI for different directions of polarization vector of incident photon. The insets represent the corresponding graphs in the absence of SOI.}
\end{figure}
\begin{figure}
\centerline{\includegraphics[width=8cm]{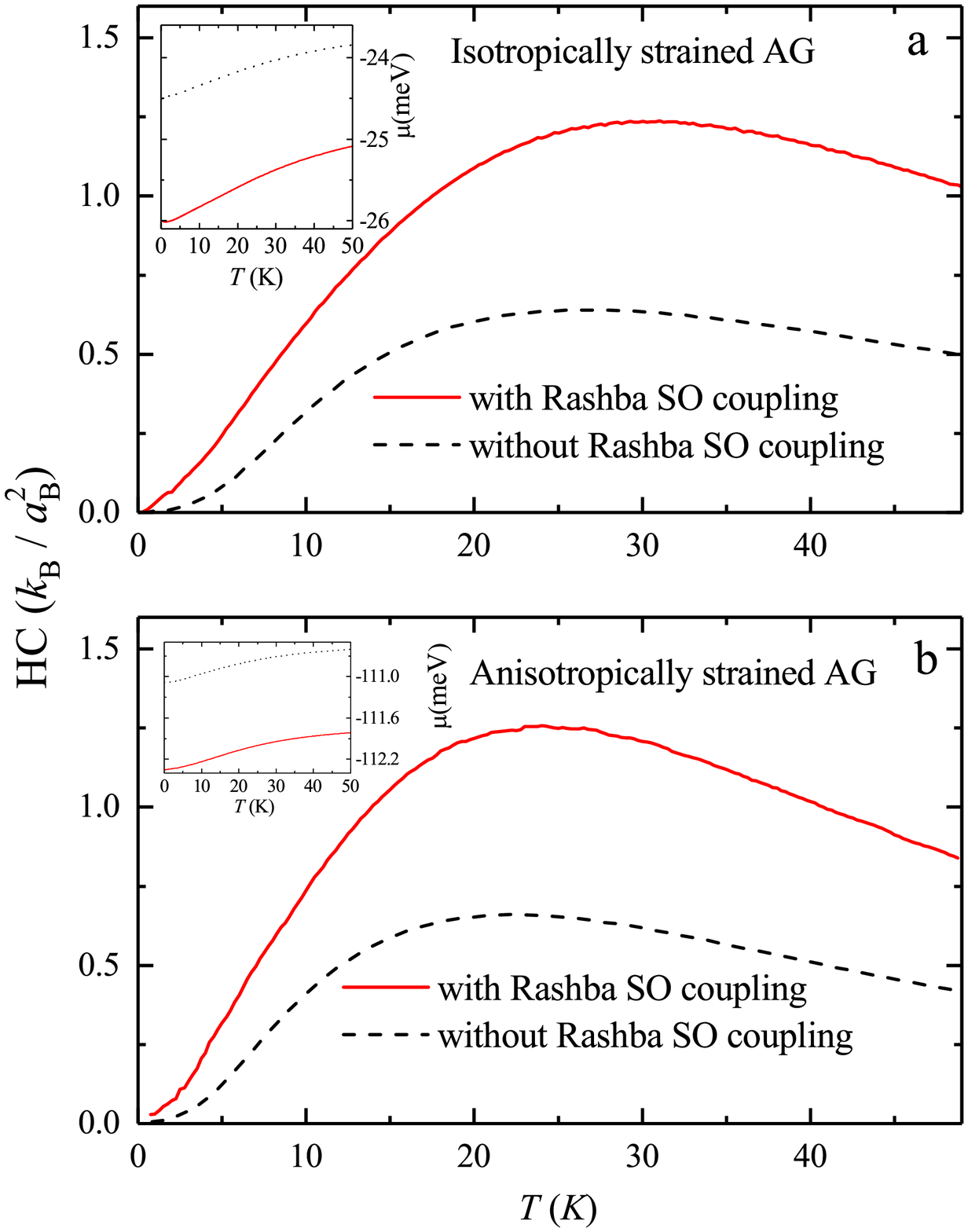}} \caption{(Colour on-line)
Heat capacity of isotropically (a) and anisotropically (b) strained AG with (red solid lines) and without (black dashed lines) Rashba SOI. The insets show the corresponding dependencies of chemical potential on temperature.}
\end{figure}

Fig.1 represents the electronic dispersion surfaces without (a) and with (b and c) Rashba SOI for isotropically strained AG. The vicinity of the $K^{\prime}$ point is mentioned by a dashed rectangle in Fig.1b, while the zoom of the corresponding region is shown in Fig.1c. It is obvious from the comparison of Fig.1a and Fig1.b, each surface splits in two due to SOI. Moreover, in the zoom of the vicinity of the $K^{\prime}$ point one can observe an obvious multiplication of Dirac points (see Fig.1c). Namely, around each Dirac point which is in the corner of the FBZ (the red line in Fig.1c) three extra Dirac-like points appear were the minibands are attached. These points are shifted from the corner of the FBZ along the diagonals of the three hexagons with the same corner and compose an equilateral triangle $K_{1}K_{2}K_{3}$ (Fig.1c). There are also three Dirac-like points of touching for each couple of the splitted surfaces which we refer as points $S_{1}$, $S_{2}$ and $S_{3}$. These points compose two equilateral triangles which are rotated by $180^{o}$ with respect to the triangle $K_{1}K_{2}K_{3}$ around the energy axis passing through the $K^{\prime}$ point (the red line in Fig.1c).

Fig.2 represents the electronic dispersion surfaces in the presence of Rashba SOI for anisotropically strained AG. Fig.2b shows the zoom of the region mentioned by the dashed rectangle in Fig.2a, while Fig.2c represents the top view of the dispersion surface of the first miniband. From Fig.2b the effect of the strain anisotropy on the symmetry of the dispersion surfaces is obvious. One can observe that the dispersion surfaces in the vicinity of $K$ points coincide with those in the vicinity of $K^{\prime}$ points when rotated by $180^{o}$ (Fig.2a). Importantly, both the Dirac points and the Dirac-like points are shifted from the corners of the FBZ (Figs.2b and c). More detailed examination shows that the shift of $K_{1}$ is along one of the axes of the FBZ, while the shifts of $K_{2}$ and $K_{3}$ are no longer along corresponding diagonals. In addition, the shift of $K_{1}$ is significantly larger than the shifts of two other Dirac-like points (Fig.2c). As a result the dispersion surfaces are neither of hexagonal nor of square symmetry and they keep only the symmetry of reflection with respect to $k_{x}$ and $k_{y}$ axes.

The DOS in the presence (red lines) and the absence (black lines) of Rashba SOI is plotted in Fig.3 for isotropically (Fig.3a) and anisotropically (Fig.3b) strained AG (in the figure $a_{B}\approx 3.57$nm is the effective Bohr radius in InAs). The results for AG without SOI are taken from the Ref. \cite{Mughnetsyan1} for comparison. An obvious multiplication of the maxima of DOS is observed. Namely, for an isotropically strained AG (Fig.3a) two peaks are replaced by eight. Each of these peaks correspond to the energy when the gradient of one of splitted surfaces is zero. The comparison of the Fig.3a with Fig.3b shows that each peak of the DOS is duplicated because of the strain anisotropy. The splittings of the left and the right peaks are very weak (less then $0.05$meV) because they correspond to the zero gradient regions of dispersion surfaces which are very close to the center of FBZ.

The effect of the miniband splitting and the symmetry change of the dispersion surfaces on the AC of AG is presented in Fig.4. The dependencies of the AC on incident photon energy for four different values of the angle $\varphi$ between the light polarization vector and the $x$ axis are shown for isotropically (Fig.4a) and anisotropically (Fig.4b) strained AG. The insets correspond to the absorption spectrum due to the transitions between the first and the second minibands in the absence of the Rashba SOI. It is noteworthy, that when SOI is absent the absorption curves corresponding to different light polarizations almost coincide for isotropically strained AG (see the inset of Fig.4a). This effect is connected with the hexagonal symmetry (symmetry of the SL) of the section of dispersion surfaces by the plane of constant energy corresponding to the allowed optical transitions in momentum space. However the strain anisotropy removes the above mentioned symmetry leading to the significant splitting of the curves which intersect at fixed value of incident photon energy (see the inset of Fig.4b). Further, one can observe pronounced maxima of AC in the presence of SOI (in both the Fig.4a and Fig.4b), which are associated to the corresponding peaks of the DOS. An obvious splitting of the curves corresponding to different polarizations of incident photon caused by SOI is also observed for both isotropically and anisotropically strained AG. Like the anisotropically strained SL without SOI (the inset of Fig.4b) there is a certain value of incident photon energy (mentioned by an arrow in Figs.4a and b) at which the values of the AC for different light polarizations coinside.

In Fig.5 the temperature dependency of HC of the AG is presented with and without Rashba SOI  for two considered different cases of strain. The insets of Fig.5 show the temperature dependency of the chemical potential assuming that the 2D concentration of electrons in conduction band is constant and are defined by the Fermi level which is on the touching point between two couples of splitted minibands. An obvious increase can be observed for the chemical potential which has larger values when there is no SOI. As we can see from the figures, HC has a non-monotonic behavior. It is zero at $T=0$, because of the vanishing density of states at the Fermi energy. With the increase of the temperature the HC initially increases due to occupation of the states in the energy regions where the density of states has maxima. However with further increase of the temperature the states with higher density in the energy scale become saturated, leading to a smaller increase of the system mean energy. As a result the HC starts to decrease from some value of $T$. Comparison of the Figs.5 a and b shows that the strain anisotropy leads to the shift of the maximum of the HC to the region of lower temperatures. On the other hand, the Rashba SOI always increases the value of the HC because it removes the twofold spin degeneracy of minibands leading to a necessity of extra energy for occupation of splitted minibands.

\section{Conclusion}
In summary, we have considered the effect of Rashba spin-orbit interaction on the energy dispersion, density of states, absorption coefficient and the heat capacity of the artificial graphene composed by InAs/GaAs QDs, taking into account the anisotropic elastic strain field due to the lattice mismatch between the materials of quantum dots and the matrix. Splitting of Dirac points due to the spin-orbit interaction has been observed. The Dirac-like points $K_{1}$, $K_{2}$ and $K_{3}$ are shifted from the $K$ point along the diagonals of adjacent hexagons in $k$-space when an isotropic strain is considered. However, in the case of anisotropic strain only $K_{1}$ point is shifted along a diagonal of the Brillouin zone. The density of states in the presence of spin-orbit interaction has eight characteristic peaks. Moreover, each of these peaks are duplicated due to the strain anisotropy. Additionally, it is shown that both the Rashba spin-orbit interaction and the strain anisotropy have qualitative effect on the measurable quantities of artificial graphene, like absorption coefficient and heat capacity. In particular, the splitting of the absorption spectra for different polarizations of incident photon, as well as the significant change in the heat capacity make the Rashba coupling an effective tool for controlling the optical and the thermal characteristics of artificial graphene.

\section{Acknowledgments}
This work was supported by the State Committee of Science of RA (research project 18T-1C223). V.M. acknowledges partial financial support from EU H2020 RISE project CoExAN (Grant No. H2020-644076). D.L. acknowledges partial financial support from Centers of excellence with BASAL/CONICYT financing, Grant FB0807, CEDENNA.

\begin{widetext}
\appendix
\section{Derivation of the Fourier transform of hydrostatic strain in two dimensional superlattice}

It is well known that in an elastic media the displacement at position $\vec{r}$ in the $l$ direction due to the unite point force applied at the origin of coordinates in $n$ direction can be expressed by Green's tensor $G_{ln}(\vec{r})$ which satisfies to the following equation \cite{Lifshits}:
\begin{equation}
\label{Green}
    \lambda_{iklm}\frac{\partial G_{ln}(\vec{r})}{\partial x_{k} \partial x_{m}}=-\delta(\vec{r})\delta_{i,n},
\end{equation}
where $\lambda_{iklm}$ is the tensor of elastic moduli.
Making the following Fourier transformations in (\ref{Green}):
\begin{equation}
\label{Fourier}
G_{ln}(\vec{r})=\int \tilde{G}_{ln}(\vec{\xi})\exp({i\vec{\xi}\vec{r}})d^{3}\xi, \hspace{1cm}
\delta(\vec{r})=(2\pi)^{-3}\int\exp({i\vec{\xi}\vec{r}})d^{3}\xi
\end{equation}
one arrives to the following equation for the Green's function Fourier transform:
\begin{equation}
\label{EqFourier}
    \sum_{klm} \lambda_{iklm} \xi_{k} \xi_{m} \tilde{G}_{ln}(\vec{\xi})=(2\pi)^{-3}\delta_{in}.
\end{equation}
In particular, for materials with cubic crystal structure
$\lambda_{iklm}=C_{12}\delta_{ik}\delta_{lm}+C_{44}(\delta_{il}\delta_{mk}+\delta_{im}\delta_{kl})+C_{an}\sum^{3}_{p=1}\delta_{ip}\delta_{kp}\delta_{lp}\delta_{mp}$, where $C_{an}=C_{11}-C_{12}-2C_{44}$ is the parameter of anisotropy. For this case after simple mathematical manipulations one gets from Eq. (\ref{EqFourier}) the following expression:
\begin{equation}
\label{DotProduct}
    (\vec{\xi} \tilde{G})_{n} \equiv \sum^{3}_{l=1} \xi_{l}\tilde{G}_{ln}(\vec{\xi})=\frac{1}{(2\pi)^{3}}\frac{\xi_{n}}{C_{44}\xi^{2}+C_{an}\xi^{2}_{n}}
    \Bigg(1+(C_{12}+C_{44})\sum^{3}_{p=1}\frac{\xi^{2}_{p}}{C_{44}\xi^{2}+C_{an}\xi^{2}_{p}}\Bigg)^{-1}.
\end{equation}

In the framework of the method of inclusions \cite{Eshelby} the ``$i$" component of the displacement caused of the existence of a single QD is as follows:
\begin{equation}
\label{Inclusions}
    D^{s}_{i}(\vec{r})= D^{s}_{i}\chi_{QD}(\vec{r})+
    \sum_{n,k}\int G_{i,n}(\vec{r}-\vec{r'})\sigma^{s}_{nk}dS'_{k},
 \end{equation}
 where $\sigma^{s}_{nk}=\sum_{pr} \lambda_{nkpr}\varepsilon^{s}_{pr}$ is the initial stress tensor, $\varepsilon^{s}_{pr}$ and $D^{s}_{i}$ are the initial strain tensor component and the initial displacement due to the lattice mismatch between the QD and the surrounded material, $\chi_{QD}(\vec{r})$ is the so called QD shape function which is 1 inside the QD and is 0 outside it. The superscript ``$s$" indicates that the expression refers to a single QD. The integration in (\ref{Inclusions}) is carried out over the surface of the QD.
 Inserting (\ref{Inclusions}) in the definition of the strain tensor:
 \begin{equation}
 \label{Strain}
     \varepsilon^{s}_{ij}=\frac{1}{2}\Big(\frac{\partial D^{s}_{i}(\vec{r})}{\partial x_{j}}+\frac{\partial D^{s}_{j}(\vec{r})}{\partial x_{i}}\Big),
 \end{equation}
and implying the Gauss's theorem one obtains:
\begin{equation}
\label{StrainGreen}
 \varepsilon^{s}_{ij}(\vec{r})=\varepsilon^{s}_{ij} \chi_{QD}(\vec{r})
 +\frac{1}{2} \sum_{nkpr} \int \Bigg[\frac{\partial^{2} G_{in}(\vec{r}-\vec{r'})}{\partial x_{j}\partial x_{k}}+\frac{\partial^{2} G_{in}(\vec{r}-\vec{r'})}{\partial x_{j}\partial x_{k}}\Bigg]\lambda_{nkpr}\varepsilon^{s}_{pr}\chi_{QD}(\vec{r'})d^{3}r',
\end{equation}
where integration is carried out over the whole 3D space.
Applying the operator $\mathfrak{F}$ of the inverse Fourier transformation to both sides of Eq. (\ref{StrainGreen}) and taking into account the convolution theorem according which $\mathfrak{F}(\int P(\vec{r}-\vec{r'})Q(\vec{r'})d\vec{r'})=
(2\pi)^{3}\mathfrak{F}(P(\vec{r}))\mathfrak{F}(Q(\vec{r}))$ for the functions $P(\vec{r})$ and $Q(\vec{r})$,
we arrive to the following expression of the strain tensor Fourier transform $\epsilon^{s}_{ij}(\vec{\xi})$ for a single QD in an elastic media:
\begin{equation}
\label{StrainFourier}
    \tilde{\varepsilon}^{s}_{ij}(\vec{\xi})= \varepsilon^{s}_{ij} \tilde{\chi}_{QD}(\vec{\xi})-
    \frac{(2 \pi)^{3}}{2} \sum_{nkpr}\Big(\xi_{i}\tilde{G}_{jn}(\vec{\xi})+\xi_{j}\tilde{G}_{in}(\vec{\xi})\Big)
   \tilde{\chi}_{QD}(\vec{\xi})\lambda_{nkpr}\xi_{k}\varepsilon^{s}_{pr},
\end{equation}
where $\tilde{\chi}_{QD}(\vec{\xi})$ is the Fourier transform of the shape function and $\vec{\xi}$ is the position vector in the inverse space. Taking into account that for cubic crystals the initial strain tensor $\varepsilon^{s}_{ij}=\varepsilon_{0}\delta_{ij}$, it is not hard to obtain from Eq. (\ref{StrainFourier}) the following expression:
\begin{equation}
\label{StrainFourierCubic}
    \tilde{\varepsilon}^{s}_{ij}(\vec{\xi})=\varepsilon_{0}\tilde{\chi}_{QD}(\vec{\xi})
    \Bigg(\delta_{ij}-\frac{(2\pi)^{3}}{2}(C_{11}+2C_{12})
    \Big[\xi_{i}(\vec{\xi} \tilde{G})_{j}+\xi_{j}(\vec{\xi} \tilde{G})_{i}\Big]\Bigg).
\end{equation}
Substituting the dot products $(\vec{\xi} \tilde{G})$ in Eq. (\ref{StrainFourierCubic}) by their corresponding expressions presented in (\ref{DotProduct}) we arrive to an analytic expression for the Fourier transform of the strain tensor:
\begin{equation}
\label{StrainFourierCubic1}
    \tilde{\varepsilon}^{s}_{ij}=\varepsilon_{0}\tilde{\chi}_{QD}(\vec{\xi})
    \Bigg(\delta_{ij}-\frac{1}{2}\frac{(C_{11}+2C_{12})\xi_{i}\xi_{j}/\xi^{2}}{1+(C_{12}+C_{44})\sum^{3}_{p=1}\frac{\xi^{2}_{p}}{C_{44}\xi^{2}+C_{an}\xi^{2}_{p}}}
    \Big[\frac{1}{C_{44}+C_{an}\xi^{2}_{i}/\xi^{2}}+\frac{1}{C_{44}+C_{an}\xi^{2}_{j}/\xi^{2}}\Big]\Bigg).
\end{equation}

Because of the linearity of the elasticity problem the strain tensor components for a one-layer QD SL is as follows:
\begin{equation}
    \label{2DSL_Strain}
    \varepsilon_{ij}(\vec{r})=\sum_{\vec{R}}\varepsilon^{s}(\vec{r}-\vec{R})=
   \sum_{\vec{R}}\int\epsilon^{s}(\vec{\xi})\exp{(\vec{\xi}(\vec{r}-\vec{R}))},
\end{equation}
where $\vec{R}$ runs over the in-plane cite vectors of QDs. On the other hand, the Fourier expansion of $\varepsilon_{ij}(\vec{r})$ has the following form:
\begin{equation}
\label{2DSL_StrainFourier}
    \varepsilon_{ij}(\vec{r})=\sum_{\vec{G}}\exp({i\vec{G}\vec{\rho}})
    \int^{\infty}_{-\infty}\tilde{\varepsilon}(\vec{G},\xi_{z})\exp({i\xi_{z}z})d\xi_{z},
\end{equation}
where $\vec{G}$ runs over the vectors of 2D reciprocal lattice and $\vec{\rho}$ is the in-plane position vector. Comparison of (\ref{2DSL_Strain}) and (\ref{2DSL_StrainFourier}) leads to the expression for the strain tensor Fourier transform for a one-layer QD SL:
\begin{equation}
\label{StrainSLbySingle}
 \tilde{\varepsilon}_{ij}(\vec{G},\xi_{z})=\frac{(2\pi)^{2}}{S_{0}}\tilde{\varepsilon}^{s}_{ij}(\vec{G},\xi_{z}).
\end{equation}
The hydrostatic strain is defined as the trace of the strain tensor:
\begin{equation}
\label{HydDef}
    \varepsilon_{h}=\sum^{3}_{i=1}\varepsilon_{ii}.
\end{equation}
Acting on the both sides of Eq. (\ref{HydDef}) by the operator $\mathfrak{F}$ and applying Eqs. (\ref{StrainFourierCubic1}) and (\ref{StrainSLbySingle}) we finally arrive to Eq. (\ref{HS}) of the Sec. II.

\end{widetext}

\end{document}